\documentclass[aps,prd,floatfix,nofootinbib,twocolumn]{revtex4}

\usepackage{graphicx}
\usepackage{amsmath}
\usepackage{amssymb}
\usepackage{subfigure}
\newcommand{\beq}{\begin{equation}}
\newcommand{\eeq}{\end{equation}}
\newcommand{\bea}{\begin{eqnarray}}
\newcommand{\eea}{\end{eqnarray}}
\newcommand{\MeV}{\;\text{MeV}}

\begin{document}
\title{ Fate of separate chiral transitions at finite $\mu_I$ under the influence of mismatched vector interactions}

\author{Zhao~Zhang}\email{zhaozhang@pku.org.cn} \author{Hai-Peng Su}
\affiliation{ School of Mathematics and Physics, North China 
Electric Power University, Beijing 102206, China}

\begin{abstract}

The flavor-mixing induced by the mismatched vector-isoscalar and vector-isovector interactions at finite baryon 
chemical potential $\mu$ and isospin chemical potential $\mu_I$ is demonstrated in the Nambu-Jona-Lasinio (NJL) 
type model of QCD. The influence of this non-anomaly flavor-mixing on the possible separate chiral transitions 
at nonzero $\mu_I$ is studied under the assumption of the effective restoration of the $U(1)_A$ symmetry. We 
find that for the weak isospin asymmetry, the two separate phase boundaries found previously can be converted 
into one only if the vector-isovector coupling $g_v^v$ is significantly stronger than the vector-isoscalar one 
$g_v^s$ without the axial anomaly. When the weak Kabayashi-Maskawa-'t Hooft (KMT) interaction is included, we 
find that the separation of the chiral transition with two critical endpoints for the relatively strong isospin 
asymmetry can still be removed owning to the vector interactions. In this case, it is not the vector coupling 
difference but the strength of $g_v^v$ which is crucial for the only phase boundary. We also point out that, in 
the NJL-type model with mismatched vector interactions, the recently proposed equivalence for chiral transitions 
at finite $\mu$ and $\mu_I$ does not hold even at the mean field approximation.

\vspace{10pt} PACS number(s): 12.38.Aw; 11.30.RD; 12.38.Lg;
\end{abstract}

\date{\today}
\maketitle

\section{Introduction}

The QCD phase diagram at finite temperature $T$ and quark chemical potential $\mu$ has attracted growing
interests. Especially, a chiral critical endpoint is predicted by some models which separates the first-order
transition and the smooth crossover. Such a point may locate at relatively higher $T$ and lower $\mu$ and
hence is promising to be explored in heavy ion collisions.  Experimently, the search for the critical endpoint
is ongoing at RHIC (BES) \cite{Mohanty:2011nm} and will be performed in the future facilities in GSI (FAIR)
and JINR (NICA). Theoretically, the existence and location of the critical endpoint are still under debate
because of the limitation of the lattice QCD computations at finite $\mu$ \cite{Philipsen:2012nu} and the
lack of other reliable theoretical methods for the non-perturbative dense QCD. Moreover, unconventional multiple
critical endpoints for the chiral transition are also proposed by some authors using effective theories
of QCD: It is found in \cite{Klein:2003fy, Toublan:2003tt} that the finite isospin chemical potential
$\mu_I$ may lead to the separate chiral transitions with two critical endpoints; when considering the
color superconductivity (CSC), the low-temperature chiral critical endpoint(s) may appear due to the
interplay between the chiral and diquark condensates
\cite{Kitazawa:2002bc,Hatsuda:2006ps,Zhang:2008wx,Zhang:2009mk,Kunihiro:2010vh}.

It is generally expected that the $U(1)_A$ anomaly may impact the QCD phase transition significantly
\cite{Pisarski:1983ms}. This point has been confirmed in model studies or Ginzberg-Landau analyses, where
the $U(1)_A$ anomaly is usually incorporated by introducing the Kabayashi-Maskawa-'t Hooft (KMT) 
interaction \cite{Kobayashi:1970ji, 't Hooft:1976fv}. The KMT interaction explicitly breaks the $U(1)_A$ 
symmetry and gives rise to the flavor-mixing among light quarks: In the
chiral symmetry breaking phase, the u quark mass may contain contributions from both d and s quark condensates;
for moderate or high baryon density, the diquark condensate for u-d pairing may make contribution to the s
quark mass. As consequence, the $U(1)_A$ anomaly may affect not only the properties of the
traditional critical endpoint \cite{Bratovic:2012qs}, but also the fate of the unconventional one:
the two critical endpoints due to nonzero $\mu_I$ \cite{Klein:2003fy, Toublan:2003tt} may be removed
by the anomaly flavor-mixing \cite{Frank:2003ve}; a new low-temperature critical endpoint could be induced
in the presence of the color flavor locking (CFL) CSC \cite{Hatsuda:2006ps}.

Even the $U(1)_A$ anomaly may influence the QCD phase transition in a nontrivial way, it's effect
could be suppressed significantly near the phase boundary. The recent lattice calculations
indicate that the $U(1)_A$ symmetry may be restored obviously near and above $T_c$ for zero $\mu$
\cite{Bazavov:2012qja,Cossu:2013uua}. The effective restoration of the $U(1)_A$ symmetry would
influence the universality class and critical properties of the chiral transition \cite{Aoki:2012yj}.
Namely, the standard analysis assuming the symmetry restoration pattern from
$SU(2)_L{\otimes}SU(2)_R{\otimes}U(1)_V$ to $SU(2)_V{\otimes}U(1)_V$ \cite{Pisarski:1983ms} can not
be applied directly. Phenomenologically, the chiral model study suggests that the location and even
the existence of the conventional critical endpoint are quite sensitive to the degree of the $U(1)_A$ 
symmetry restoration \cite{Bratovic:2012qs}. If the anomaly related flavor-mixing is very weak
near the phase boundary, the low-temperature critical endpoint due to the CFL phase \cite{Hatsuda:2006ps}
may be directly ruled out; in contrast, the two critical endpoints due to the isospin asymmetry
\cite{Klein:2003fy, Toublan:2003tt} could be still possible because of the decouple of light quarks.

Nevertheless, it is also probable that the non-anomaly flavor-mixing of light quarks can be induced by other
ingredients of QCD, especially under some condition. The main purpose of this paper is to study the possible
non-anomaly flavor-mixing and its effect on the chiral phase transition at finite temperature and density
under the isospin asymmetry. In particular, we will concentrate on the fate of the two critical endpoints 
due to the separate chiral transitions found in \cite{Klein:2003fy, Toublan:2003tt} with the assumption of 
the effective restoration of the $U(1)_A$ symmetry near the phase boundary. In addition, the validity of 
the recently proposed phase quenching in the mean field approximation (MFA) of QCD-models at finite $\mu$ 
and $\mu_I$ \cite{Hanada:2011ju,Hanada:2012es} will be checked by taking into account the non-anomaly 
flavor-mixing.

Our starting point is the four-quark vector interactions with different coupling strengths in the isovector
and isoscalar channels. In the literature, the role of vector interactions on the chiral transition has been
extensively studied in the NJL-type model of QCD. A well-known result is that the chiral transition at finite
$\mu$ is weakened by the vector-isoscalar interaction $g_v^s(\bar{\psi}\gamma_\mu\psi)^2$ and it would be no
critical point for strong $g_v^s$
\cite{Asakawa:1989,Kitazawa:2002bc,Fukushima:2008is,Zhang:2009mk,Bratovic:2012qs}. When considering the
two-flavor CSC, it is found that the $g_v^s$ in a proper range can lead to the new low-temperature
critical endpoint(s) \cite{Kitazawa:2002bc}, especially under the constraint of electric-charge
neutrality \cite{Zhang:2009mk, Kunihiro:2010vh}
\footnote{Note that under the electric-charge neutrality, the electric chemical potential influences the
chiral phase transition in the similar way as the isoscalar vector interaction \cite{Zhang:2008wx}.}.
The sufficiently strong $g_v^s$ is also used to account for the shrinkage of the first-order region in the quark mass
plane \cite{Fukushima:2008is}. On the other hand, the chromomagnetic instability for the two-flavor neutral CSC is
found to be suppressed by the vector-isovector interaction $g_v^v(\bar{\psi}\vec{\tau}\gamma_{\mu}\psi)^2$
\cite{Zhang:2009mk}. It is also demonstrated in \cite{Kunihiro:1991qu,Sasaki:2006ws,Ferroni:2010xf} that both
vector interactions play important role in reproducing the lattice flavor diagonal and off-diagonal
susceptibilities at finite $T$ and $\mu$. In general, the $g_v^v$ and $g_v^s$ are independent coupling constants
under the condition of chiral symmetry. As far as we know, the non-anomaly flavor-mixing due to
the mismatched vector interactions in the vacuum has been discussed in Ref.~\cite{Takizawa:1990ay} based on a
three-flavor NJL model. However, the effect of the possible non-anomaly flavor-mixing at finite temperature and
density is ignored in all the previous studies, which will be investigated in this paper.

The main part of this work can be regarded as an extension of Ref.~\cite{Frank:2003ve} by including the
vector interactions, where the roles of the non-anomaly flavor-mixing and the vector-isovector
interaction are focused on with the assumption of the effective suppression of the $U(1)_A$ anomaly.
The paper is organized as follows. In Sec.II, the extended NJL model with the vector interactions is
introduced and the arguments for the vector coupling difference are given. In Sec.III, we show that
the non-anomaly flavor-mixing arises at nonzero $\mu$ and $\mu_I$ due to the mismatched vector 
interactions. The effects of the vector interactions on the separation of the chiral transition
and the validity of the phase quenching at finite $\mu$ and $\mu_I$ are demonstrated in Sec.IV.
In Sec.V, we discuss and summarize.

\section{ Extended NJL-type model with mismatched vector interactions }

\subsection{ The general four-quark interaction model with mismatched vector interactions under the chiral symmetry }

We start with the following Lagrangian of four-quark interaction model for two-flavor QCD
\begin{eqnarray}
\mathcal{L}^{(4)}=\mathcal{L}^{(4)}_{\text{sym}}+\mathcal{L}^{(4)}_{\text{det}}
, \label{eq:lagr}
\end{eqnarray}
with
 \begin{align}
 \mathcal{L}^{(4)}_{\text{sym}}&=g_{s1}\sum_{a=0}^{3}[(\bar{\psi}\tau^{a}\psi)^2+(\bar{\psi}\tau^{a}i\gamma_5\psi)^2]\nonumber\\
&-g_{v2}\sum_{a=0}^{3}[(\bar{\psi}\tau^{a}\gamma_{\mu}\psi)^2+(\bar{\psi}\tau^{a}\gamma_{\mu}\gamma_5\psi)^2]\nonumber\\
&-g_{v3}[(\bar{\psi}\tau^{0}\gamma_{\mu}\psi)^2+(\bar{\psi}\tau^{0}\gamma_{\mu}\gamma_5\psi)^2]\nonumber\\
&-g_{v4}[(\bar{\psi}\tau^{0}\gamma_{\mu}\psi)^2-(\bar{\psi}\tau^{0}\gamma_{\mu}\gamma_5\psi)^2]
 \label{eq:sym}
\end{align}
and
\begin{align}
\mathcal{L}^{(4)}_{\text{det}}&=g_{s2}\{det[\bar{\psi}(1-\gamma_5)\psi]+h.c.\}\nonumber\\
&=g_{s2}[(\bar{\psi}\psi)^2-(\bar{\psi}\vec{\tau}\psi)^2-(\bar{\psi}i\gamma_5\psi)^2+(\bar{\psi}\vec{\tau}i\gamma_5\psi)^2]
\label{eq:det},
\end{align}
where $\tau^{0}$, and $\vec{\tau}$ refer to the unit matrix and Pauli matrices in the flavor space,
respectively.
The former term $\mathcal{L}^{(4)}_{\text{sym}}$ in \eqref{eq:lagr} is the general Fierz-invariant
form of the four-quark interactions in color-singlet channels which respecting the global flavor symmetries
of $SU(2)_V{\otimes}SU(2)_A{\otimes}U(1)_V{\otimes}U(1)_A$~\cite{Klevansky:1992}. The latter one
$\mathcal{L}^{(4)}_{\text{det}}$ is the KMT interaction induced by the gauge configurations of instanton and 
anti-instanton \cite{'t Hooft:1976fv},  which only possesses the $SU(2)_V{\otimes}SU(2)_A{\otimes}U(1)_V$ 
global flavor symmetries.

As mentioned, we will focus on the flavor-mixing arising from the mismatched vector interactions at finite density.
We see that three of the four independent coupling constants in $\mathcal{L}^{(4)}_{\text{sym}}$ are related 
to the vector and axial vector interactions. Generally, the nozero sum $g_{v3}+g_{v4}$ implies
that the vector coupling strength in the isovector channel is different from that in the isoscalar one.
Namely, these two coupling constants are independent of each other under the chiral symmetry.
Similarly, the non-vanishing $g_{v3}-g_{v4}$ indicates the mismatched axial-vector interactions in the isovector
and isoscalar channels. How the vector coupling difference gives rise to the non-anomaly flavor-mixing at
finite density will be detailed in next section.

Since we mainly study the chiral phase transition in the MFA, the axial-vector interactions in Lagrangian 
\eqref{eq:lagr} will be ignored
\footnote{The axial-vector interaction may be responsible for the deviation of the chiral
magnetic effect in the recent lattice calculation compared to the analytic formula, as
proposed in Ref.~\cite{Zhang:2012rv}.}.
Hence in our calculations,  we only consider the the following effective Lagrangian

\begin{align}
 \mathcal{L}^{(4)}_{\text{eff}}&=g_{s1}\sum_{a=0}^{3}[(\bar{\psi}\tau^{a}\psi)^2+(\bar{\psi}\tau^{a}i\gamma_5\psi)^2]\nonumber\\
&+g_{s2}[(\bar{\psi}\psi)^2-(\bar{\psi}\vec{\tau}\psi)^2-(\bar{\psi}i\gamma_5\psi)^2+(\bar{\psi}\vec{\tau}i\gamma_5\psi)^2]\nonumber\\
&-g_v^s(\bar{\psi}\gamma_{\mu}\psi)^2-g_v^v(\bar{\psi}\vec{\tau}\gamma_{\mu}\psi)^2,
\label{eq:eff}
\end{align}
where the independent coupling constants are reduced to four.

\subsection{ Unequal vector coupling constants in the mean field Hartree-Fork approximation }

Here we stress that the vector coupling difference in the MFA can also arise from a very popular version of the NJL
model \cite{Klevansky:1992}
\begin{align}
 \mathcal{L}^{(4)}_{\text{}}&=g_{s1}\sum_{a=0}^{3}[(\bar{\psi}\tau^{a}\psi)^2+(\bar{\psi}\tau^{a}i\gamma_5\psi)^2]\nonumber\\
&+g_{s2}[(\bar{\psi}\psi)^2-(\bar{\psi}\vec{\tau}\psi)^2-(\bar{\psi}i\gamma_5\psi)^2+(\bar{\psi}\vec{\tau}i\gamma_5\psi)^2]\nonumber\\
&-g_v\sum_{a=0}^{3}[(\bar{\psi}\tau^{a}\gamma_{\mu}\psi)^2+(\bar{\psi}\tau^{a}\gamma_{\mu}\gamma_5\psi)^2],
\label{eq:VNJL}
\end{align}
in which only one vector coupling $g_v$ is adopted. In the Hartree approximation, there is no difference
between the coupling strengths of the two vector interactions at the mean field level for Lagrangian \eqref{eq:VNJL}.

However, the effective vector couplings (in the sense of direct interaction) in the isoscalar and isovector 
channels will differ from each other if the Fock contribution is also considered. 
For a four-fermion interaction, the Fock contribution can be 
easily evaluated according to its Fierz transformation \cite{Klevansky:1992}. Taking into account the exchange terms, 
the effective direct four-quark interactions of the Lagrangian \eqref{eq:VNJL} take the following form:
\begin{align}
\mathcal{L}^{(4)}&_{\text{eff-direct}} =\mathcal{L}^{(4)}+\mathcal{L}^{(4)}_{\text{Fock}}\nonumber\\
  &=(g_{s1}+g_{s2}+\frac{g_{s2}}{2N_c})[(\bar{\psi}\psi)^2+(\bar{\psi}i\gamma_5\vec{\tau}\psi)^2]\nonumber\\
&+(g_{s1}-g_{s2}-\frac{g_{s2}}{2N_c})[(\bar{\psi}\vec{\tau}\psi)^2+(\bar{\psi}i\gamma_5\psi)^2]\nonumber\\
&-g_v\sum_{a=0}^{3}[(\bar{\psi}\tau^{a}\gamma_{\mu}\psi)^2+(\bar{\psi}\tau^{a}\gamma_{\mu}\gamma_5\psi)^2]\nonumber\\
&-(\frac{g_v}{N_c}+\frac{1}{2}\frac{g_{s1}}{N_c})(\bar{\psi}\tau^{0}\gamma_{\mu}\psi)^2\nonumber\\
&-(\frac{g_v}{N_c}-\frac{1}{2}\frac{g_{s1}}{N_c})(\bar{\psi}\tau^{0}\gamma_{\mu}\gamma_5\psi)^2,
 \label{eq:HatreeFock}
\end{align}
where $N_c$ is the color number of the quarks. The effective Lagrangian \eqref{eq:HatreeFock} clearly
shows that the exchange terms give rise to the vector coupling difference in the Hartree-Fock approximation (HFA), 
which is at the order of $O(1/N_c)$ compared to $g_{s1}$ and $g_v$. Note that the similar result in a three-flavor 
NJL model has been given in \cite{Takizawa:1990ay}, where the influence of the induced non-anomaly
flavor-mixing on the spin content of the nucleon is investigated.

If both $g_v$ and $g_{s1}$ in \eqref{eq:VNJL} originate from the color current-current interaction
 $g(\bar{\psi}\gamma_\mu\lambda^a_c\psi)^2$, they fulfill the relation $g_v=g_{s1}/2$ according to the Fierz 
transformation. In this case, the vector coupling difference shown in \eqref{eq:HatreeFock} becomes
\beq
\delta{g_v}=g_v^s-g_v^v=\frac{2}{N_c}g_v=\frac{g_{s1}}{N_c}.\label{eq:diffoge}
\eeq
This equation indicates that the $g_v^s$ in \eqref{eq:eff} may be larger than the $g_v^v$ and their difference
is considerable compared to $g_v$ or $g_{s1}$ for $N_c=3$. However, there is no coupling strength difference 
between the isoscalar and isovector axial-vector interactions in the HFA for this situation.

\subsection{ Constraints on the vector interactions from the lattice chiral curvatures }

Even Eq.~\eqref{eq:diffoge} implies that the coupling $g_v^v$ is weaker than the $g_v^s$, it is also 
possible that the $g_v^v$ may be stronger than the $g_v^s$. This can be understood from the curvature difference 
for the chiral phase transition at finite baryon and isospin chemical potentials obtained in recent lattice
calculations \cite{Cea:2012ev}.

For small baryon and/or isospin densities, the chemical potential dependence of the pseudo-critical temperature 
for the chiral crossover can be expressed as
\beq
T_c(\mu_q,\mu_i)=T_c+A_q\mu_q^2+B_i\mu_i^2+\textit{O}(\mu_{q/i}^4,\mu_q^2\mu_i^2)\,,\\
\label{eq:Tc}
\eeq
where $T_c$ is the chiral pseudo-critical temperature at zero quark chemical potential
(In this subsection, $\mu_q$ and $\mu_i$ are used to refer to the quark baryon and isospin chemical
potentials, respectively).
Notice that $T_c(\mu_q,\mu_i)$ is an even function of $\mu_{q/i}$ \cite{D'Elia:2009tm}. So at the order
of $\mu_{q/i}^2$,  we can expand $T_c(\mu_{q/i}^2)$ as
\beq
T_c(\mu_{q/i}^2)=T_c(1-\kappa_{q/i}\frac{\mu_{q/i}^2}{T_c^2}),
\eeq
where the two chiral curvatures are defined as
\beq
\kappa_{q/i}=-T_c\frac{dT_c(\mu^2)}{d\mu_{q/i}^2}|_{\mu=0}.
\eeq
The lattice QCD simulation in \cite{Cea:2012ev} suggests that the curvature $\kappa_q$ is about
$10\%$ greater than $\kappa_i$.

Recently, the role of the vector coupling strength $g_v^s$ on the determination of $\kappa_q$ has been studied
in a Polyakov-loop enhanced three-flavor NJL model \cite{Bratovic:2012qs}. It is found that the $\kappa_q$ decreases
with the $g_v^s$ and to reproduce the lattice $\kappa_q$ in this model the $g_v^s$ must keep relatively larger
value compared to the scalar coupling strength $g_s$. The authors of Ref.~\cite{Bratovic:2012qs} then propose the lattice
$\kappa_q$ can be used as a useful constraint on the strength of $g_v^s$. 

We can directly extend this idea  to determine the $\kappa_i$  by replacing $\mu_q$ with $\mu_i$. 
As will be demonstrated in the next section, the coupling $g_v^v$ influences the curvature $\kappa_i$ in 
the similar way as the $g_v^s$ does on the $\kappa_q$.
In particular, the $\kappa_i$ and $\kappa_q$ obtained at the MFA of the two-flavor NJL model will take the same
value for $g_v^v=g_v^s$. In other words, the lattice curvature difference between the $\kappa_i$ and
$\kappa_q$ can be regarded as an useful evidence for the unequal vector coupling strengths.

Since the two-flavor lattice calculation in \cite{Cea:2012ev} indicates that the $\kappa_i$ is less than the $\kappa_q$, 
we thus infer that the $g_v^v$ may be larger than the $g_v^s$ near the chiral phase boundary for zero and small quark
chemical potential. Following the spirit of Ref.~\cite{Bratovic:2012qs}, our numerical study suggests that the $g_v^v$ 
is about $10\%$ larger than the $g_v^s$ near $T_c$ according to the lattice curvatures in \cite{Cea:2012ev}.
This conclusion is quite different from the estimation given in \eqref{eq:diffoge}.

\subsection{ Constraints on the vector interactions from the couplings of vector mesons to nucleons and lattice susceptibilities }

In Ref.~\cite{Sasaki:2006ws}, it is argued that the ratio of the couplings of $\omega$ and $\rho$ mesons to
nucleons can be used as a constraint on the vector coupling difference. In the chirally broken phase, the empirical
value for this ratio is given by $g_{\omega NN}/g_{\rho NN}\simeq {3}$, whereas in the chirally symmetric phase it
is expected to be one. It is then proposed that the ratio $g_v^v/g_v^s$ is located in the range from $1/3$ to 1. 
The study in \cite{Sasaki:2006ws} also suggests that the flavor off-diagonal susceptibility for vanishing 
chemical potential is very sensitive to the vector coupling difference.

In addition, another quite similar estimation is given in Ref.~\cite{Ferroni:2010xf}, where the vector coupling 
difference is expressed as the function of two susceptibilities $\chi_q$ and $\chi_I$ under some assumptions. 
Using the lattice data for these susceptibilities as input, it is found that the $g_v^v$ is always less than the 
$g_v^s$: their difference is quite large below $T_c$ which approaches zero rapidly above $T_c$ for zero chemical 
potential.

All the arguments given in the above subsections suggest that the vector interactions are repulsive
(namely, $g_v^s$ and $g_v^v$ are all positive), but the relation between the $g_v^s$ and $g_v^v$ remains uncertain. 
Usually, the two vector couplings in the NJL model must take almost the same strength to reproduce the
vacuum masses of the $\omega$ and $\rho$ mesons at the MFA. However, the description of vector meson properties within the
 NJL formalism is not quite reliable since the momentum cutoff is not large enough compared to the $\omega$ and $\rho$
masses. In addition, the $T$-$\mu$ dependence of the vector couplings is also unknown. Due to these uncertainties, the 
$g_v^s$ and $g_v^v$ in the Lagrangian \eqref{eq:eff} will be treated as the free parameters in the following study.

\section{ vector-interaction induced flavor-mixing and the thermal dynamical potential at finite baryon and isospin chemical potentials}

In this section, we shall demonstrate that the vector coupling difference can lead to non-anomaly flavor-mixing at finite
baryon and isospin densities.

The full Lagrangian of two-flavor NJL model with the interaction \eqref{eq:eff} reads
\begin{eqnarray}
\mathcal{L}=\bar{\psi}\left(i\partial_{\mu}\gamma^{\mu}+\gamma_0\hat{\mu}-\hat{ m}_0\right)\psi+\mathcal{L}^{(4)}_{\text{eff}}
, \label{eq:lagrNJL}
\end{eqnarray}
where the quark chemical potentials are introduced and $\hat{m}_0 = diag(m_u,m_d)$ is the current quark matrix. We
shall adopt the isospin symmetric quark masses with $m_u =m_d \equiv m_0$. The $\hat{\mu}$ in the
Lagrangian \eqref{eq:lagrNJL} is the matrix of the quark chemical potentials which takes form
\begin{equation}
\hat{\mu}=\bigg(\begin{array}{cc}
    \mu_u & \\
     & \mu_d\end{array}
 \bigg)=\bigg(\begin{array}{cc}
    \mu-\mu_I & \\
     & \mu+\mu_I\end{array}
 \bigg),
\end{equation}
with
\begin{eqnarray}
\mu=\frac{\mu_u+\mu_d}{2}=\frac{\mu_B}{3}~~~\mathrm{and}~~~\mu_I=\frac{\mu_u-\mu_d}{2}=\frac{\delta\mu}{2}\label{eq:chemicalp}.
\end{eqnarray}
In \eqref{eq:chemicalp}, $\mu_B$ ($\mu_I$) is the baryon (isospin) chemical potential, which corresponds to the
conserved baryon (isospin) charge. In our notations, the $\mu_I$ refers to half of the difference between the 
$\mu_u$ and $\mu_d$ .

At finite densities, the quark chemical potentials are shifted by the vector interactions. Here we use $\mu'$ to 
denote the modified quark chemical potential. Note that the u quark density is different from the d quark one under 
the isospin asymmetry. Considering this point, the shifted quark chemical potentials take the form
\begin{align}
\mu'_{u(d)}&=\mu_{u(d)}-2g_v^s(\rho_{u}+\rho_{d})-2g_v^v(\rho_{u(d)}-\rho_{d(u)})\nonumber\\
&=\mu_{u(d)}-2(g_v^s+g_v^v)\rho_{u(d)}-2(g_v^s-g_v^v)\rho_{d(u)},
\label{eq:mixmu}
\end{align}
or
\bea
\mu'=\mu-2g_v^s(\rho_{u}+\rho_{d}),\quad \mu_{I}'=\mu_{I}-2g_v^v(\rho_{u}-\rho_{d}),
\label{eq:shiftmu}
\eea
where
\beq
\rho_{u(d)}=\langle{\psi_{u(d)}^\dag\psi_{u(d)}}\rangle
\eeq
is the u (d) quark number density. Eq.~\eqref{eq:mixmu} clearly shows that due to the vector coupling difference, 
not only the $\rho_{u}$ but also the $\rho_{d}$ give contribution to the effective chemical potential of u
quark, and vise versa. This implies that the flavor-mixing arises at the MFA due to the vector interaction. As 
mentioned, this mixing has nothing to do with the axial anomaly. The modified chemical potentials can also be 
rearranged as Eq.~\eqref{eq:shiftmu}, which indicates that $\mu$ and $\mu_I$ are shifted by the isoscalar and 
isovector vector interactions, respectively.

Formally, the non-anomaly flavor-mixing shown in \eqref{eq:mixmu} for the modified chemical potentials is
quite similar to the anomaly flavor-mixing for the constituent quark masses induced by the instantons, namely
\bea
M_{u(d)}=m_0-4g_{s1}\phi_{u(d)}-4g_{s2}\phi_{d(u)}\label{eq:mixm},
\eea
where
\beq
\phi_{u(d)}=\langle{\bar{\psi}_{u(d)}}\psi_{u(d)}\rangle
\eeq
is the u (d) quark condensate. So at finite $\mu_I$, the four-fermion interactions in \eqref{eq:eff} can give
rise to two types of flavor-mixing at the MFA for non-vanishing $\delta{g_v}$ and $g_{s2}$.

Using the conventional technique, the mean field thermal dynamical potential of the Lagrangian \eqref{eq:lagrNJL} at
finite temperature and chemical potentials is expressed as
\begin{align}
\Omega&(T,\mu_u,\mu_d)=\nonumber\\
&\sum_{f=u,d}\Omega_0(T,{\mu'}_f;M_f)+2g_{s1}(\phi_u^2+\phi_d^2)+4g_{s2}\phi_u\phi_d\nonumber\\
&-(g_v^s+g_v^v)(\rho_u^2+\rho_d^2)-2(g_v^s-g_v^v)\rho_u\rho_d,
\label{eq:omega}
\end{align}
where $\Omega_0(T,{\mu'}_f;M_f)$ is the contribution of a quasi-particle gas of the
flavor $f$ which takes the form
\begin{align}
\Omega_0&(T,{\mu'}_f;M_f)=\nonumber\\
&-2N_cT\int{\frac{d^3p}{(2\pi)^3}} \Big[\ln[1+\exp(-(E_f-{\mu'}_f)/T)]\nonumber\\
&+\ln[1+\exp(-(E_f+{\mu'}_f)/T)]\Big]\nonumber\\
&-2N_c\int{\frac{d^3p}{(2\pi)^3}}E_f\theta(\Lambda^2-\vec{p}^2), \label{eq:omega0}
\end{align}
with the quasi-particle energy $E_f=\sqrt{\vec{p}^2+M_f^2}$. The $\Lambda$ in \eqref{eq:omega0} is the
parameter of three-momentum cutoff in the NJL model. We see that besides the modified chemical potential
 $\mu_f'$, the flavor-mixing due to the vector coupling difference is also explicitly demonstrated in \eqref{eq:omega}
via the direct coupling between the $\rho_u$ and $\rho_d$. This is also analogous to the instanton induced flavor-mixing:
in addition to the constituent mass $M_f$, the u quark condensate couples directly to the d quark one, which is also 
shown in \eqref{eq:omega}.

Minimizing the thermal dynamical potential \eqref{eq:omega}, the motion equations for the mean fields
$\phi_u$, $\phi_d$, $\rho_u$ and $\rho_d$ are determined through the coupled equations
\begin{equation}
\frac{\partial\Omega}{\partial\phi_u}=0,\quad\frac{\partial\Omega}{\partial\phi_d}=0,
\quad\frac{\partial\Omega}{\partial\rho_u}=0,\quad\frac{\partial\Omega}{\partial{\rho_d}}=0.
\end{equation}
This set of equations is then solved for the fields $\phi_u$, $\phi_d$, $\rho_u$ and $\rho_d$ as functions
of the temperature and chemical potentials. When there exist multi roots of these coupled equations, the
solution corresponding to the minimal thermodynamical potential is favored.

\section{ chiral phase transition under the influence of mismatched vector interactions }

\begin{figure}[t]
\hspace{-.0\textwidth}
\begin{minipage}[t]{.42\textwidth}
\includegraphics*[width=\textwidth]{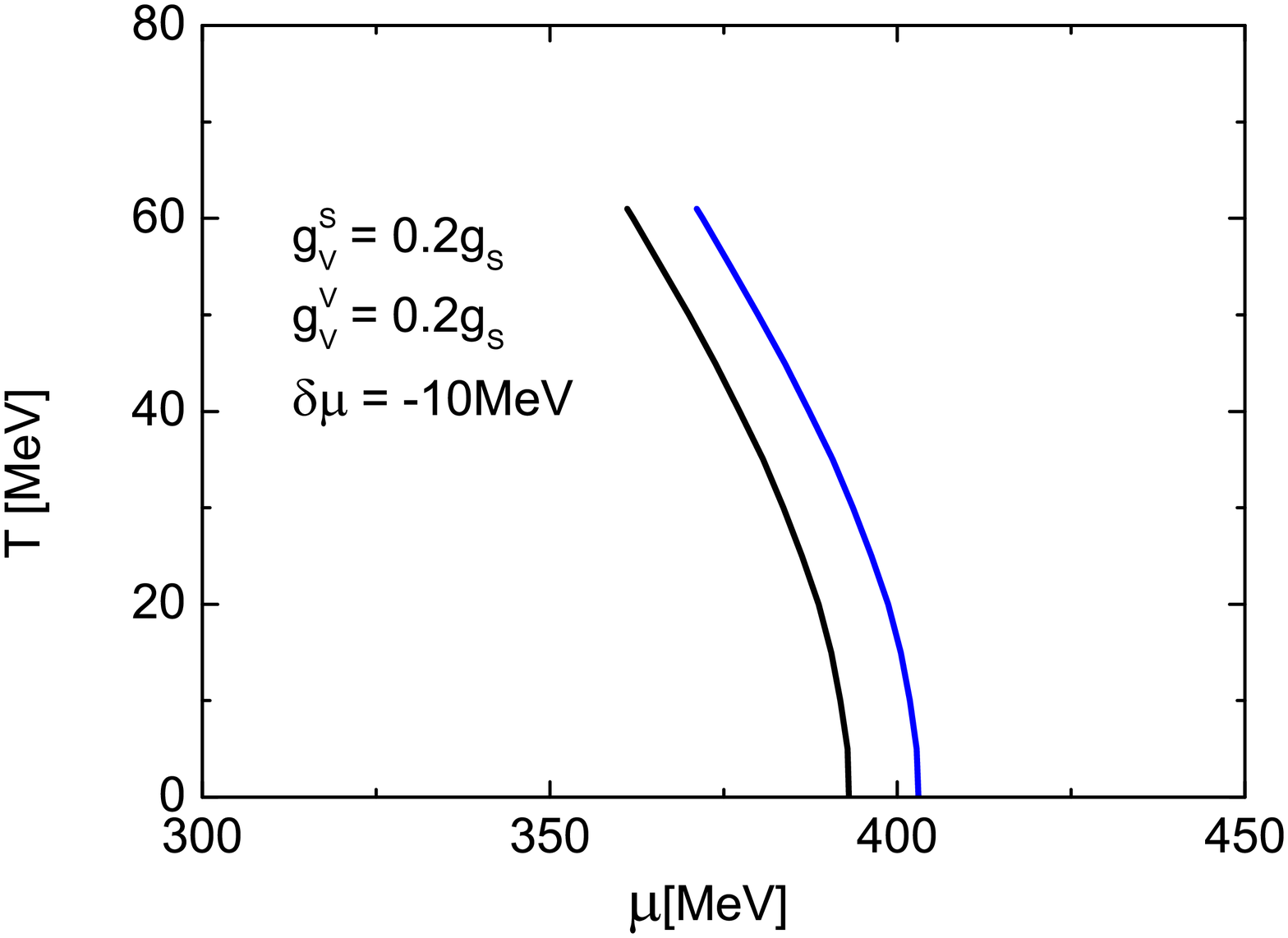}
\centerline{(a) }
\end{minipage}
\hspace{-.05\textwidth}
\begin{minipage}[t]{.42\textwidth}
\includegraphics*[width=\textwidth]{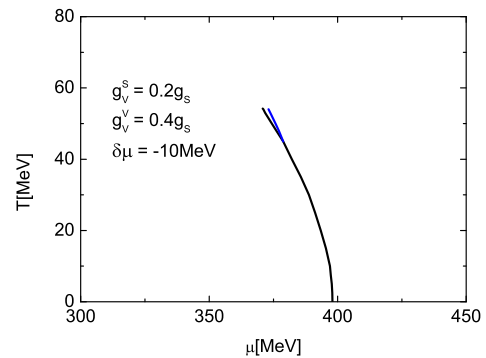}
\centerline{(b) }
\end{minipage}
\hspace{-.05\textwidth}
\begin{minipage}[t]{.42\textwidth}
\includegraphics*[width=\textwidth]{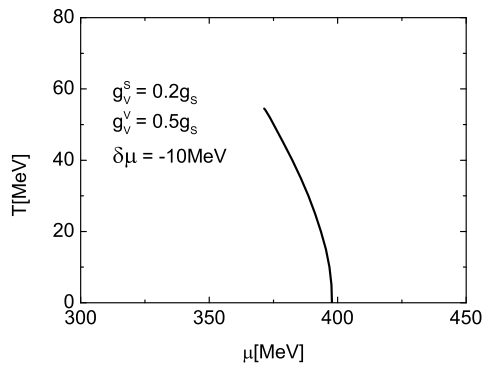}
\centerline{(c) }
\end{minipage}
\caption{The $T$-$\mu$ phase diagrams for varied vector-isovector coupling $g_v^v$ at $\delta\mu=-10 \MeV$ 
without the axial anomaly. The vector-isoscalar coupling is fixed as $g_v^s=0.2g_s$ (the $g_s$ is the scalar coupling). 
The solid line stands for the first-order chiral boundary. }
\label{fig:pdui05GVfixed}
\end{figure}

\begin{figure}
\hspace{-.0\textwidth}
\begin{minipage}[t]{.42\textwidth}
\includegraphics*[width=\textwidth]{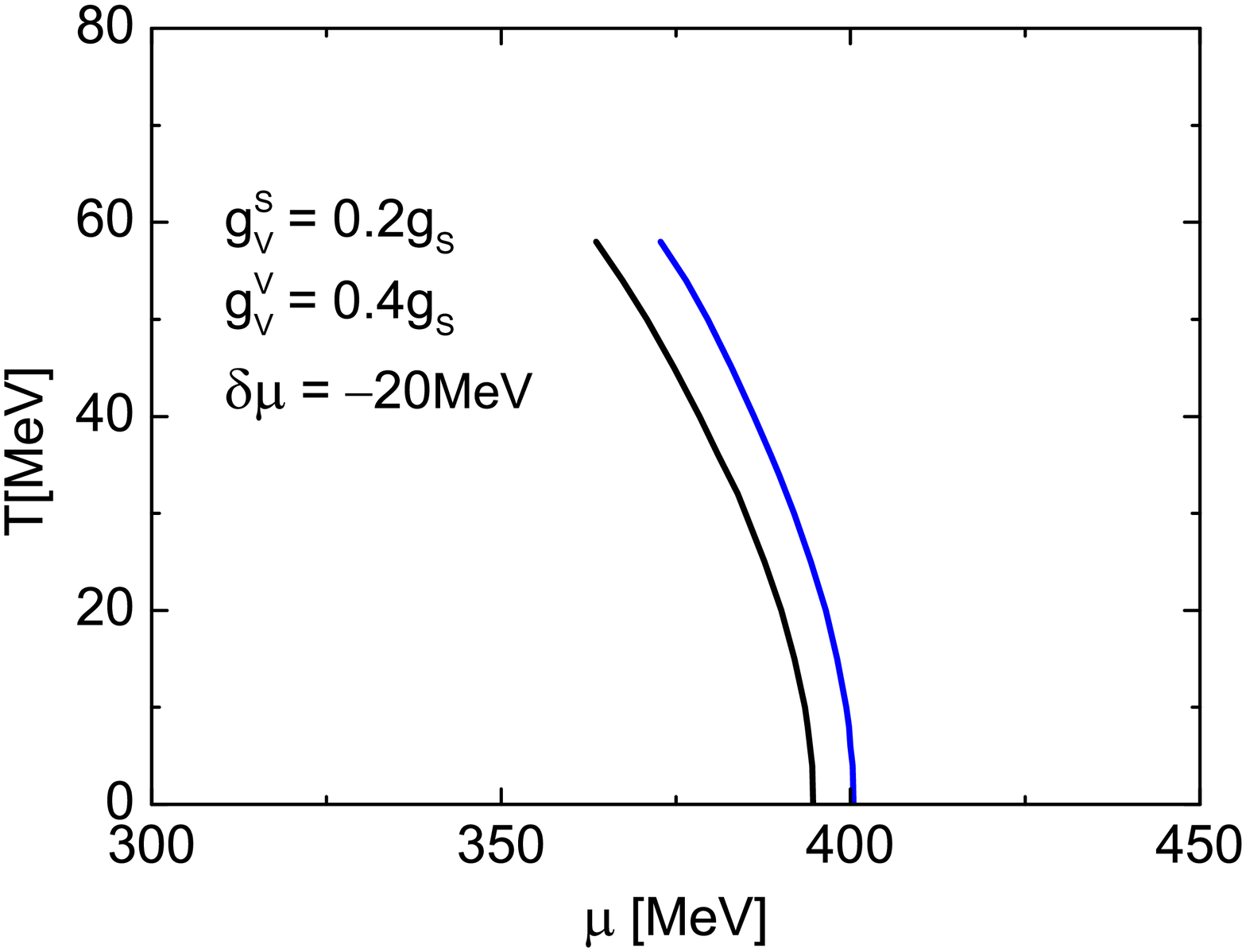}
\centerline{(a) }
\end{minipage}
\hspace{-.05\textwidth}
\begin{minipage}[t]{.42\textwidth}
\includegraphics*[width=\textwidth]{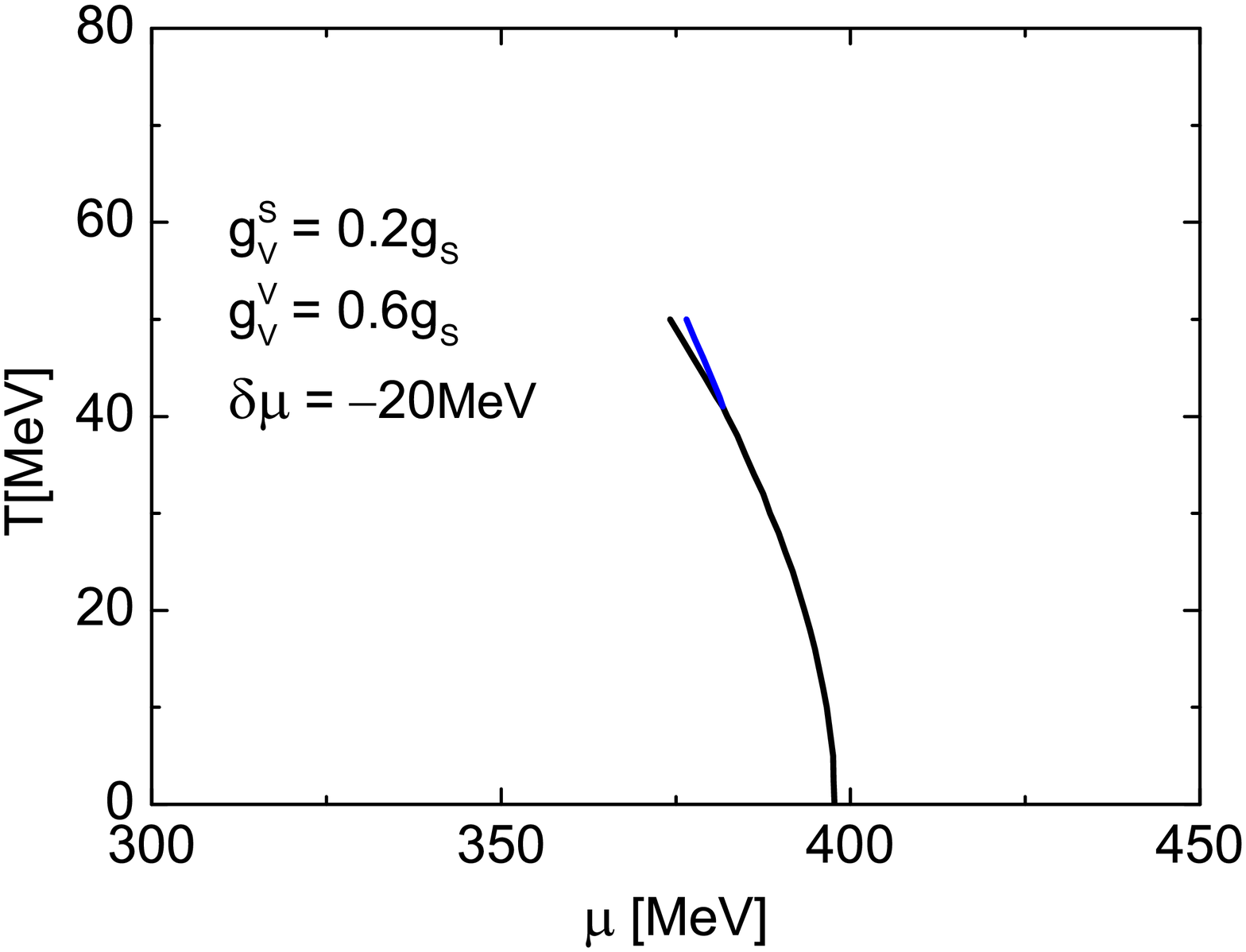}
\centerline{(b) }
\end{minipage}
\hspace{-.05\textwidth}
\begin{minipage}[t]{.42\textwidth}
\includegraphics*[width=\textwidth]{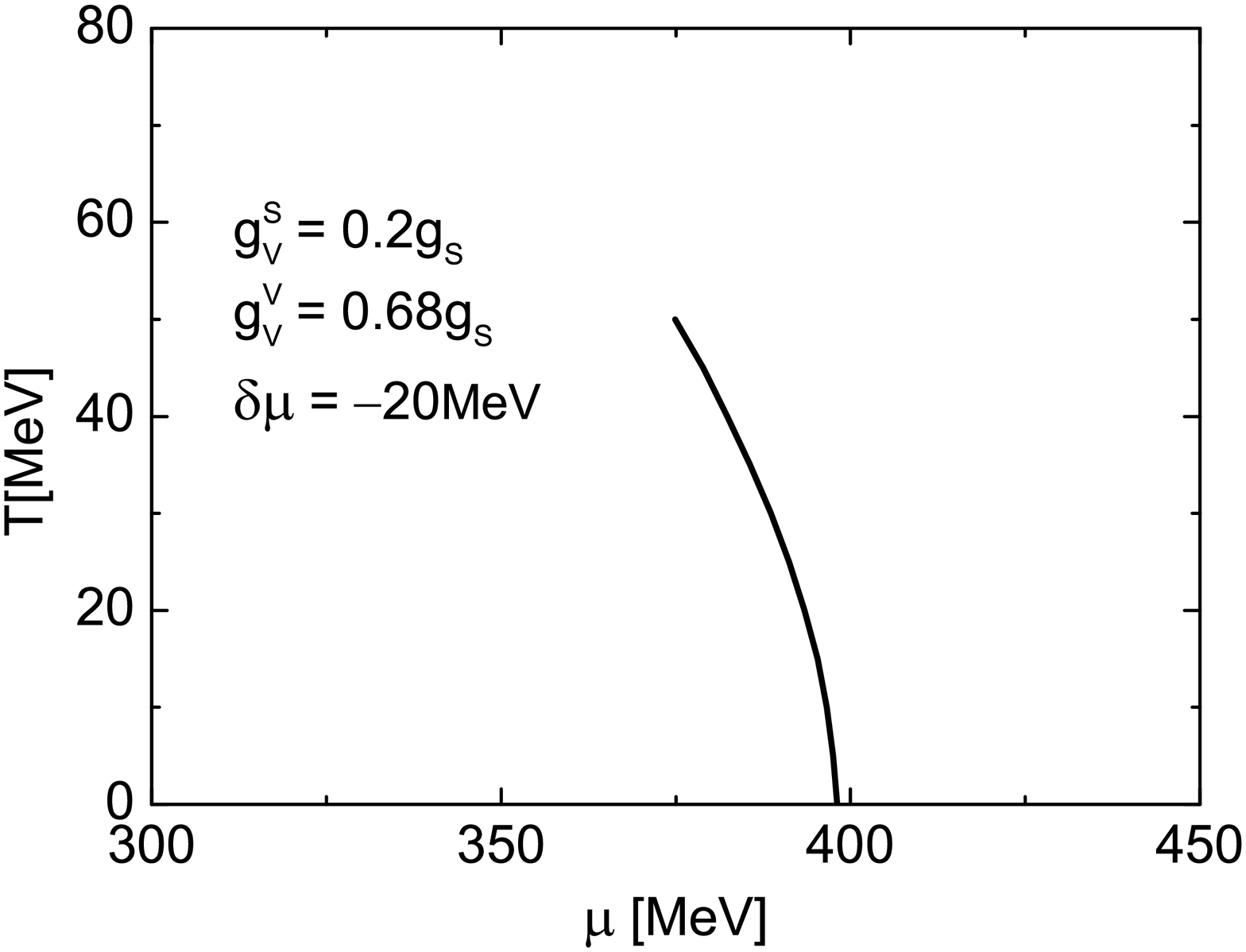}
\centerline{(c) }
\end{minipage}
\caption{The $T$-$\mu$ phase diagrams for varied vector-isovector coupling $g_v^v$ at $\delta\mu=-20 \MeV$
without the axial anomaly. The fixed vector-isoscalar coupling $g_v^s$ is the same as in Fig.~\ref{fig:pdui05GVfixed}.
The solid line stands for the first-order chiral boundary.}
\label{fig:pdui10GVS0p2fixed}
\end{figure}

\begin{figure}
\hspace{-.0\textwidth}
\begin{minipage}[t]{.42\textwidth}
\includegraphics*[width=\textwidth]{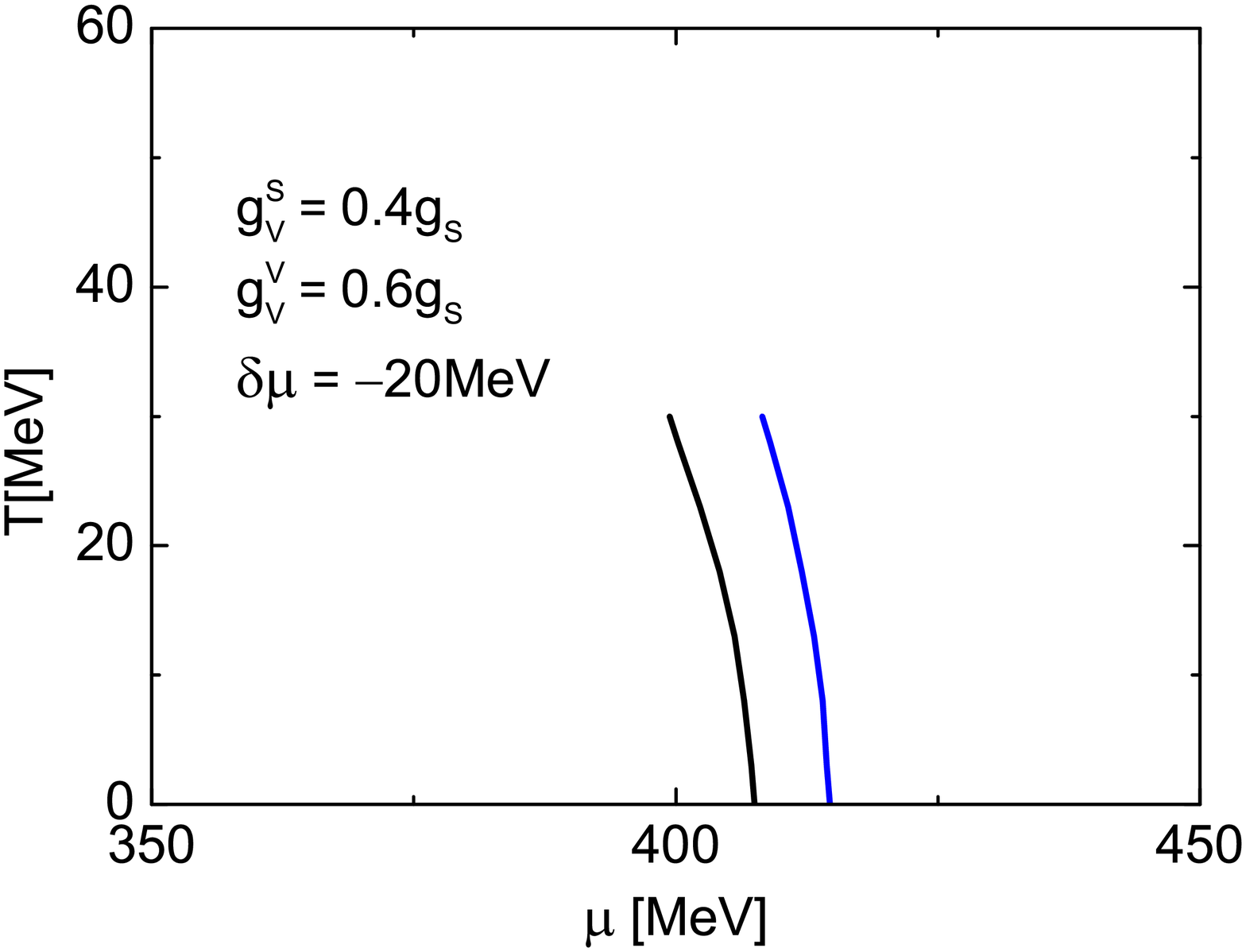}
\centerline{(a) }
\end{minipage}
\hspace{-.05\textwidth}
\begin{minipage}[t]{.42\textwidth}
\includegraphics*[width=\textwidth]{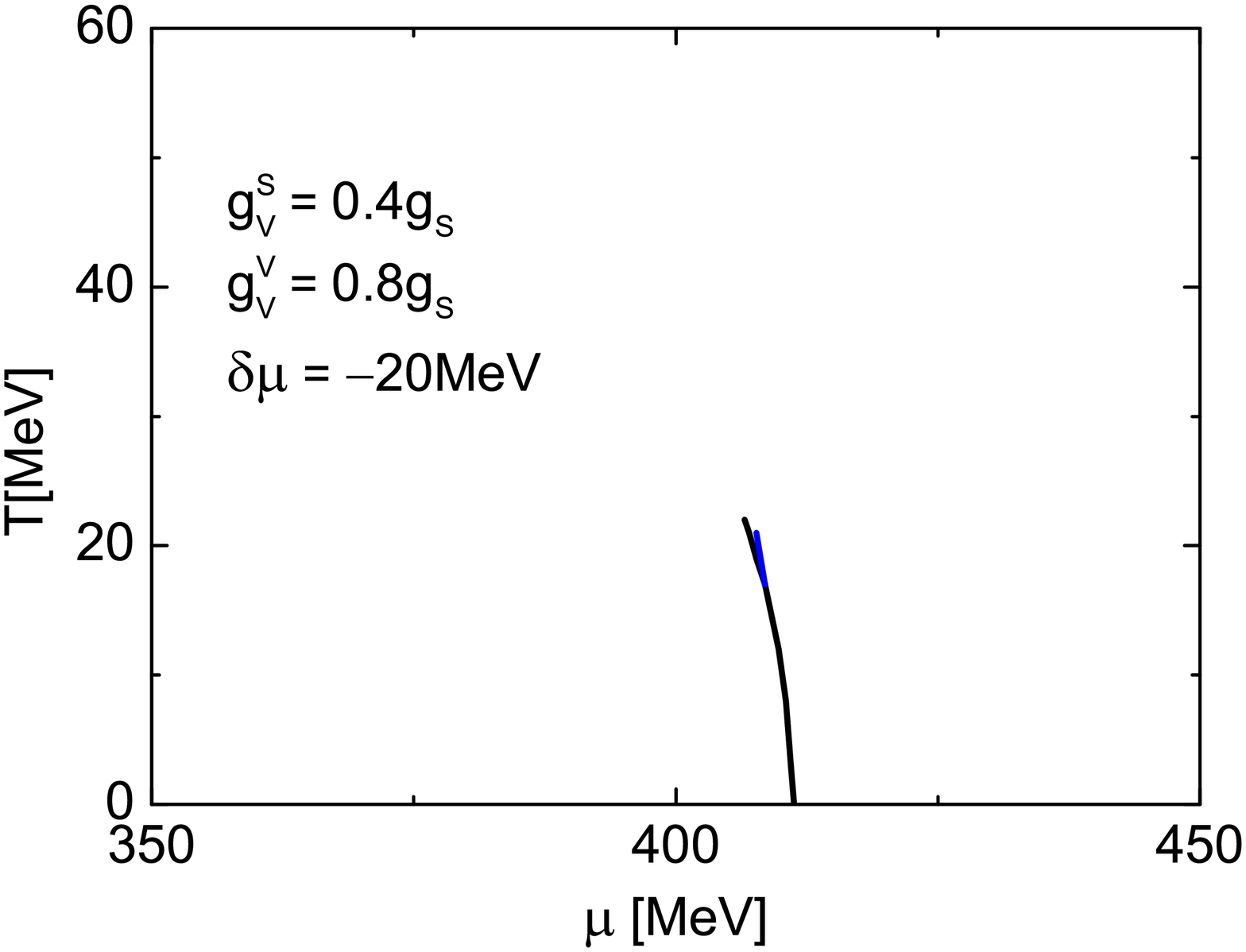}
\centerline{(b) }
\end{minipage}
\hspace{-.05\textwidth}
\begin{minipage}[t]{.42\textwidth}
\includegraphics*[width=\textwidth]{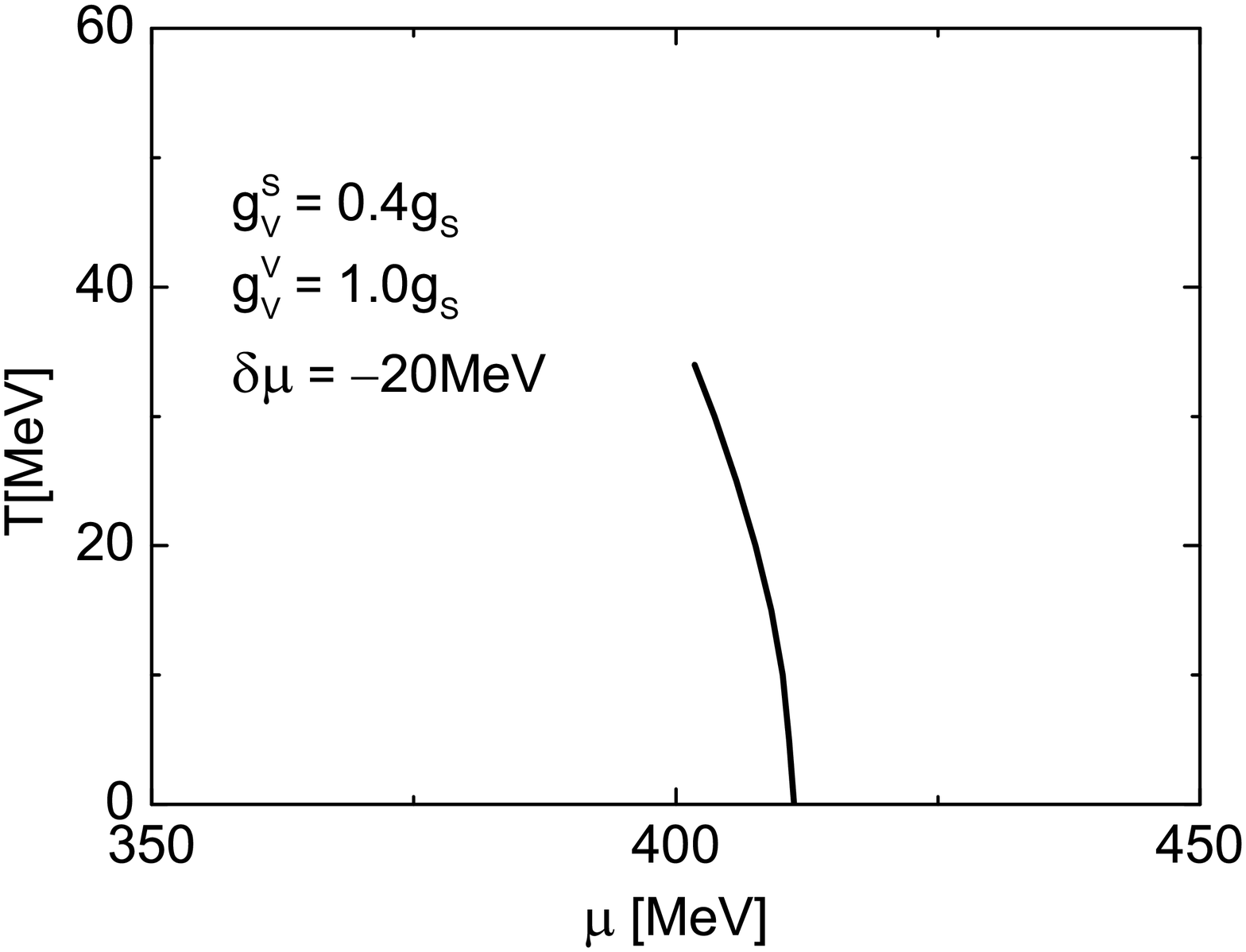}
\centerline{(c) }
\end{minipage}
\caption{The $T$-$\mu$ phase diagrams for varied vector-isovector coupling $g_v^v$ at $\delta\mu=-20 \MeV$
without the axial anomaly. The vector-isoscalar coupling is fixed as $g_v^s=0.4$ (relative to the scalar coupling $g_s$).
The solid line stands for the first-order chiral boundary.}
\label{fig:pdui10GVS0p4fixed}
\end{figure}

As mentioned, the separate chiral transitions because of finite $\mu_I$ \cite{Toublan:2003tt,Klein:2003fy} 
can be removed by the flavor-mixing induced by the axial anomaly \cite{Frank:2003ve}. Since the instanton density 
may be suppressed significantly near the phase boundary, we revisit this problem by taking into account the non-anomaly 
flavor-mixing due to the mismatched vector interactions. We shall check whether the emergence of the two critical
endpoints is sensitive to the couplings $g_v^s$ and $g_v^v$. In addition, the so called chiral equivalence at finite $\mu$ 
and $\mu_I$ is also checked in the MFA of NJL model by including the vector interactions .

For comparison, we follow the notations in Ref.~\cite{Frank:2003ve} and introduce two parameters $\alpha$ and $g_s$
which are defined as
\beq
g_{s1}=(1-\alpha)g_s, \quad\quad\quad g_{s2}=\alpha g_s.
\eeq
Here the $\alpha$ means the ratio of the KMT interaction in the scalar-pseudoscalar channel,  which is treated as a
free parameter in the following calculations. The other model parameters, namely the current quark mass $m_0$, the
scalar coupling constant $g_s$ and the three-momentum cutoff $\Lambda$ are all adopted from \cite{Frank:2003ve},
which take the values
\begin{equation}
m_0=6\ \mathrm{MeV},\quad \Lambda=0.590\ \mathrm{GeV},\quad g_s\Lambda^2=2.435.
\end{equation}
These parameters are fixed by the pion mass, the pion decay constant, and  the chiral condensate of the QCD vacuum.

\begin{figure}[t]
\begin{center}
\includegraphics[width=1.0\columnwidth]{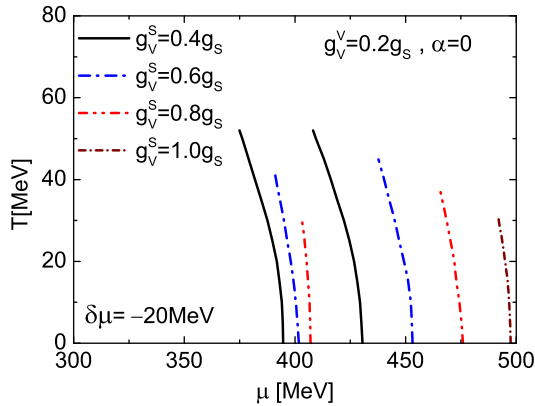} \vspace{1em}\\
\end{center}
\caption{\label{Fig:gs0p2} The $T$-$\mu$ phase diagrams for varied vector-isoscalar coupling $g_v^s$ at 
$\delta\mu=-20 \MeV$.  The vector-isovector coupling is fixed as $g_v^v=0.2$ (relative to the scalar coupling $g_s$).
The axial anomaly is ignored. All the lines stand for the first-order chiral boundaries.}
\label{fig:pdui10GVV0p2fixed}
\end{figure}

\subsection{ Fate of separate chiral transitions under the weak isospin asymmetry without the axial anomaly }

The role of the mismatched vector interactions on the separation of the chiral transition at finite 
$T$-$\mu$ under the weak isospin asymmetry is first investigated by switching off the KMT interaction. 
We focus on whether the two critical endpoints found previously could be ruled out by the non-anomaly 
flavor-mixing without the help of the axial anomaly.

We first study the cases for $g_v^v>g_v^s$ with a fixed small coupling $g_v^s=0.2g_s$ under 
the rather weak isospin asymmetry $\delta\mu=-10 \MeV$
(Note that the $\mu_I$ defined in \cite{Frank:2003ve} corresponds to the $\delta\mu$ in our notations). 
The $T$-$\mu$ phase diagrams for varied $g_v^v$  are shown in Fig.~\ref{fig:pdui05GVfixed}. 
For $g_v^v=g_v^s$, Fig.~\ref{fig:pdui05GVfixed}.(a) shows two separate first-order phase boundaries, 
which correspond to the chiral transitions for the u and d quarks, respectively. This is natural because 
of the decouple of the u and d quarks. For $g_v^v=0.4g_s$, Fig.~\ref{fig:pdui05GVfixed}.(b) shows that 
only one first-order chiral boundary emerges at the low temperature, but it splits into two lines at 
the relatively higher temperature. So there are still two critical endpoints. Further increasing $g_v^v$ 
to $0.5g_s$, Fig.~\ref{fig:pdui05GVfixed}.(c) displays that only one phase boundary appears. So we 
really observe that the two separate phase boundaries can be changed into one by the non-anomaly 
flavor-mixing induced by the mismatched vector interactions.

We then increase the isospin asymmetry to $\delta\mu=-20\MeV$ with the $g_v^s$ unchanged 
(The typical value of $\delta\mu$ in heavy ion collisions may be within this range, as estimated in \cite{Frank:2003ve}). 
We obtain the similar phase diagrams by varying the $g_v^v$, which are displayed in
Fig.~\ref{fig:pdui10GVS0p2fixed}. Compared to Fig.~\ref{fig:pdui05GVfixed}, a more large vector coupling 
difference is needed to convert the two phase boundaries into one because of the enhanced isospin asymmetry.

The above calculation for $\delta\mu=-20\MeV$ is further extended to a fixed moderate coupling 
$g_v^s=0.4g_s$. The phase diagrams for varied $g_v^v$ with $g_v^v>g_v^s$ are shown in 
Fig.~\ref{fig:pdui10GVS0p4fixed}, which is still analogous to Fig.~\ref{fig:pdui05GVfixed}.
In contrast to Fig.~\ref{fig:pdui10GVS0p2fixed}, a more strong $g_v^v$ is required for the conversion of  
the two phase transitions into one due to the enlarged $g_v^s$. Fig.~\ref{fig:pdui10GVS0p4fixed} also shows 
that the chiral transition is first softened and then strengthened with $g_v^v$. By comparison, the 
chiral transition is always weakened with the increase of $g_v^s$.

So for the weak isospin asymmetry, Figs.~\ref{fig:pdui05GVfixed}-\ref{fig:pdui10GVS0p4fixed} show that the
chiral transition separation can be removed by the mismatched vector interactions, even without the 
instanton induced flavor-mixing. Actually, all the three sets of phase diagrams in 
Figs.~\ref{fig:pdui05GVfixed}-\ref{fig:pdui10GVS0p4fixed} are quite similar to Fig.~2 in Ref.~\cite{Frank:2003ve}
obtained by changing the $\alpha$. In this sense, the non-anomaly flavor-mixing due to the vector 
coupling difference plays the similar role as the KMT interaction. 

However, Figs.~\ref{fig:pdui05GVfixed}-\ref{fig:pdui10GVS0p4fixed} indicate that the $g_v^v$ must be much 
stronger than the $g_v^s$ for turning the two chiral transitions into one: the $g_v^v$ is at least twice 
as strong as the $g_v^s$ to remove the separation.  Of course, the fate of the separate chiral transitions  
depends on not only the vector coupling difference, but also the magnitudes of $g_v^v$ and $g_v^s$. Here we 
do not show the results for $g_v^v>g_v^s$ with a fixed strong $g_v^s$ since in this case only 
crossover transition appears.

On the contrary, we don't find the coincidence of the detached  phase boundaries for $g_v^v<g_v^s$. In 
Fig.~\ref{fig:pdui10GVV0p2fixed}, we show the phase diagrams for $\delta\mu=-20 \MeV$ with varied $g_v^s$ and 
fixed coupling $g_v^v=0.2g_s$. We see that the two separate phase boundaries get farther rather than closer 
with the increase of $|\delta{g_v}|$ for $g_v^v<g_v^s$, which is quite different from what shown in
Figs.~\ref{fig:pdui05GVfixed}-\ref{fig:pdui10GVS0p4fixed}. 

The reason can be traced aback to Eqs.~\eqref{eq:mixmu} and~\eqref{eq:shiftmu}. First, according to
Eq.~\eqref{eq:shiftmu}, the $|\mu'_I|$ is explicitly less than the $|\mu_I|$ since the signs of $\mu_I$ and
$-2g_v^v(\rho_{u}-\rho_{d})$ in $\mu'_I$ are different for $g_v^v>0$. So for $g_v^v>g_v^s$ with 
a fixed $g_v^s$, increasing $g_v^v$ implies not only the enhancement of the flavor-mixing but also the reduction 
of $|\mu_I'|$ (see Fig.~\ref{fig:effuiGV04} in next subsection). This is why the two phase boundaries approach
each other with the $g_v^v$, as shown in Figs.~\ref{fig:pdui05GVfixed}-\ref{fig:pdui10GVS0p4fixed}. Second, near 
the left side of the right phase boundary, the $\rho_d$ is remarkably larger than the $\rho_u$ because of the 
significant suppression of the d quark mass; but around the left side of the left phase boundary, the difference 
between the $\rho_d$ and $\rho_u$ is relatively small. So for $g_v^s>g_v^v$, the flavor-mixing term 
$-(g_v^s-g_v^v)\rho_{d}$ in $\mu_u'$ impacts the right phase boundary more significantly in contrast to what 
the corresponding term $-(g_v^s-g_v^v)\rho_{u}$ in $\mu_d'$ does on the left phase boundary, according to
Eq.~\eqref{eq:mixmu}. This is why the right phase boundary moves more rapidly towards the higher $\mu$ with 
$g_v^s$ in contrast to the left one, as shown in Fig.~\ref{fig:pdui10GVV0p2fixed}.

If $g_v^s$ or/and $g_v^v$ are strong enough, the first-order chiral transition will change into crossover and 
it would be no critical point. Owning to the vector interactions, it is possible that one of the two phase boundaries 
first disappears while the other one still remains with the change of the vector interactions 
(In contrast, the two critical endpoints always appear at the same temperature in Ref.~\cite{Toublan:2003tt,Klein:2003fy}). 
Such a case is really observed in Fig.~\ref{fig:pdui10GVV0p2fixed} for very strong vector interaction $g_v^s=1.0g_s$. 
In the next subsection, we will show that the emergence of only one critical endpoint via this manner does not 
require very strong vector interaction when the weak KMT interaction is included.

\begin{figure}[t]
\begin{center}
\includegraphics[width=1.0\columnwidth]{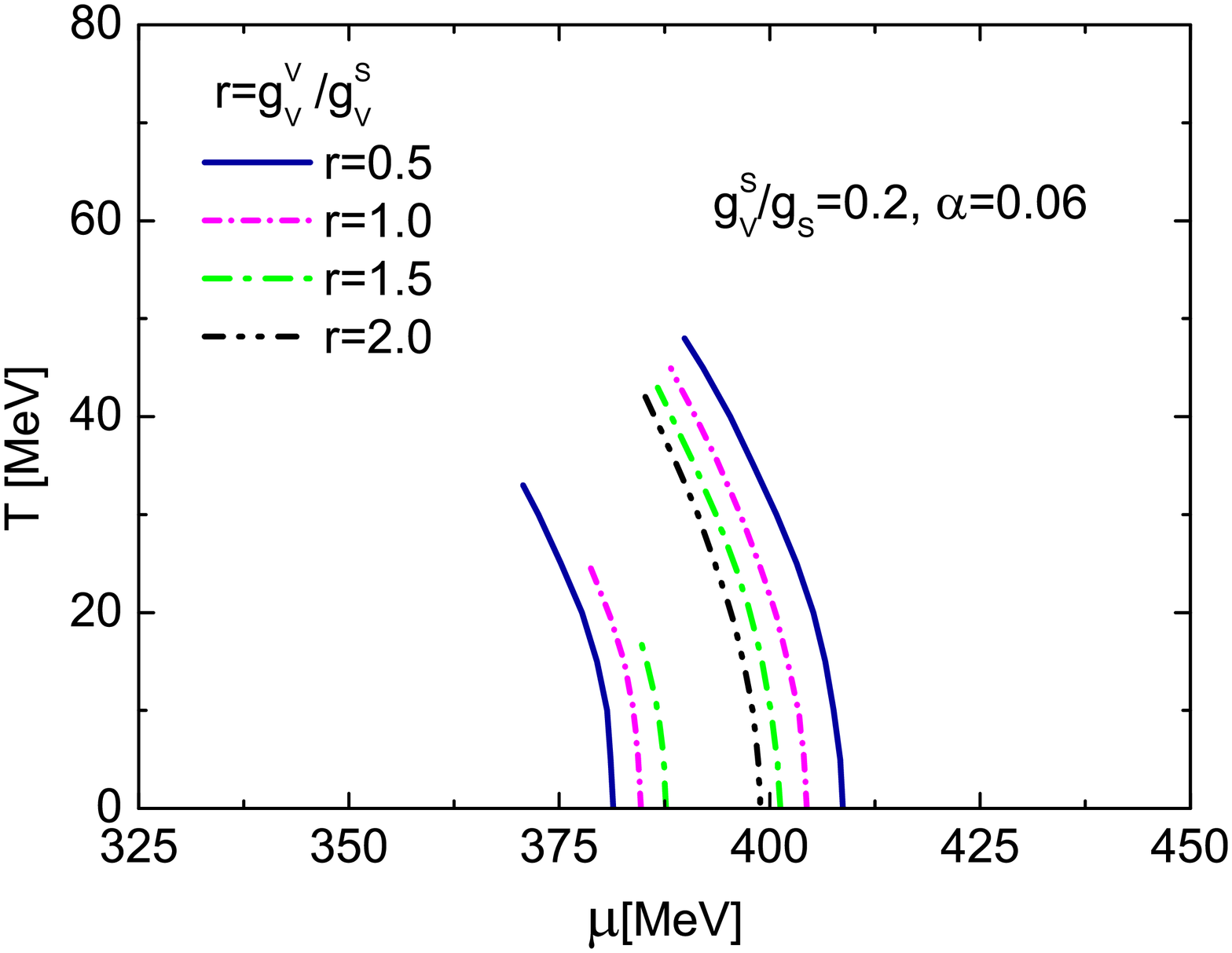} \vspace{1em}\\
\end{center}
\caption{\label{Fig:gs0p2} The first order chiral boundaries in the $T$-$\mu$ plane for varied vector-isovector coupling
 $g_v^v$ at $\delta\mu=-60 \MeV$. The parameter $\alpha$ for the KMT interaction and vector-isoscalar coupling $g_v^s$ are
fixed as 0.06 (about one third of the vacuum value) and 0.2 (relative to the scalar coupling $g_s$), respectively. }
\label{fig:pdui30GV02}
\end{figure}

\begin{figure}[t]
\begin{center}
\includegraphics[width=1.0\columnwidth]{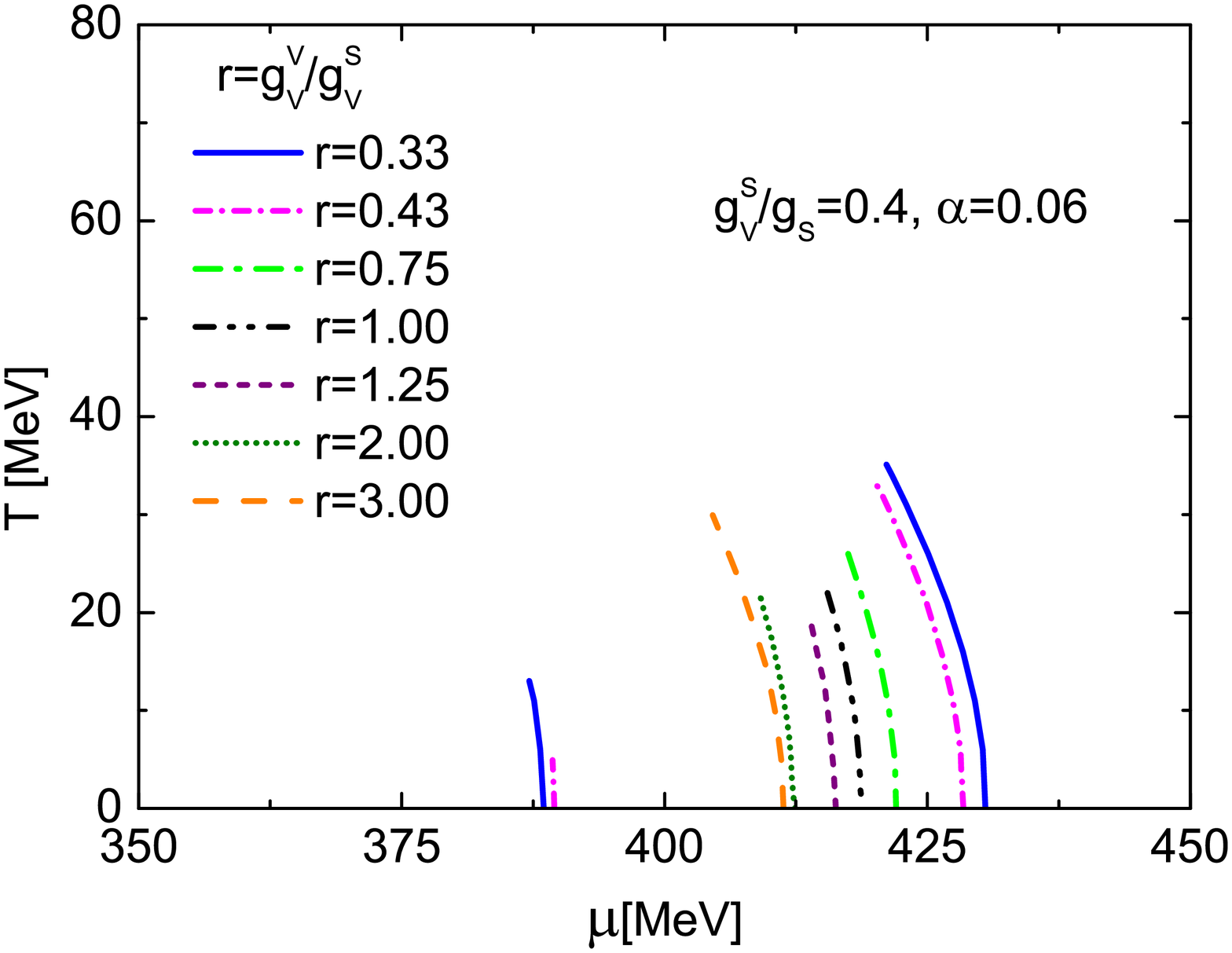} \vspace{1em}\\
\end{center}
\caption{\label{Fig:gs0p2} The first order chiral boundaries in the $T$-$\mu$ plane for varied vector-isovector coupling
 $g_v^v$ at $\delta\mu=-60 \MeV$. The parameter $\alpha$ for the KMT interaction and vector-isoscalar coupling $g_v^s$ are 
fixed as 0.06 (about one third of the vacuum value) and $0.4$ (relative to the scalar coupling $g_s$), respectively. }
\label{fig:pdui30GV04}
\end{figure}

\subsection{ Fate of separate chiral transitions at finite $\mu_I$ under the influence of both vector interactions and the axial anomaly }

\begin{figure}[t]
\begin{center}
\includegraphics[width=1.0\columnwidth]{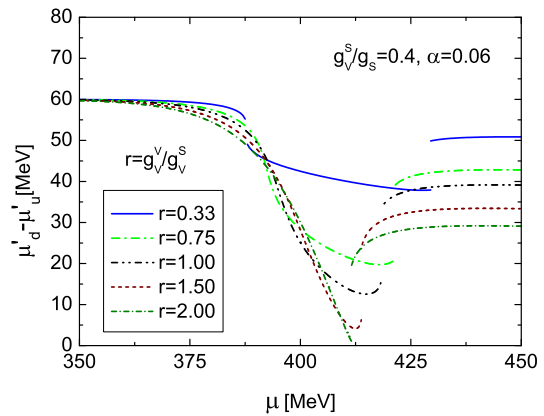} \vspace{1em}\\
\end{center}
\caption{ The difference between the effective chemical potentials of u and d quarks as a function of
$\mu$ for varied vector-isovector coupling $g_v^v$ at $\delta\mu=-60 \MeV$ and $T=10$ \MeV. The vector-isoscalar coupling $g_v^s$ and
parameter $\alpha$ for the KMT interaction are fixed as 0.4 (relative to the scalar coupling $g_s$) and 0.06 (about one third of the vacuum value), respectively. }
\label{fig:effuiGV04}
\end{figure}

\begin{figure}[t]
\begin{center}
\includegraphics[width=1.0\columnwidth]{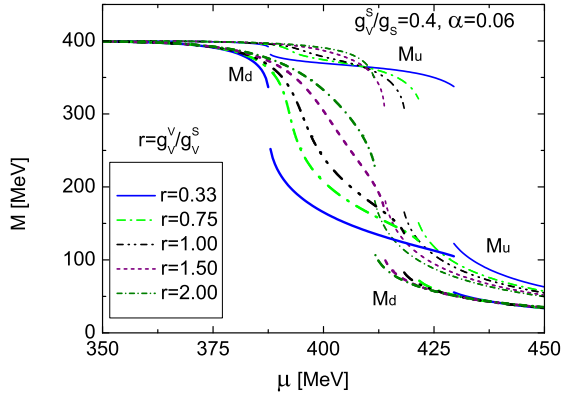} \vspace{1em}\\
\end{center}
\caption{ The u and d quark masses as functions of $\mu$ for varied vector-isovector coupling $g_v^v$ at
$\delta\mu=-60 \MeV$ and $T=10$ \MeV. The vector-isoscalar coupling $g_v^s$ and parameter $\alpha$ for the 
KMT interaction are the same as in Fig.~\ref{fig:effuiGV04}. }
\label{fig:MGV04}
\end{figure}

In Ref.~\cite{Frank:2003ve}, it is found that the separate chiral transitions at fixed $\delta\mu$=60$\MeV$
\footnote{The chiral transition with the same $\delta\mu$ is also studied in \cite{Toublan:2003tt} in a NJL type model,
where the axial anomaly is ignored.}
only appear for $\alpha<\alpha_c=0.12$, where the critical value $\alpha_c$ is argued to be less than the vacuum $\alpha$.
Here, we assume that the physical $\alpha$ near the phase boundary is obviously less than the $\alpha_c$ due to the effective
suppression of instantons. We shall concentrate on whether the two critical endpoints could still appear under the same isospin
asymmetry as in Ref.~\cite{Frank:2003ve} by including the vector interactions. For comparison, the $\delta\mu$ is fixed 
as $-60 \MeV$ in our numerical study 
(Changing the sign of $\delta\mu$ does not alter the conclusion since it only indicates the interchange of u and d quarks.).
We choose a comparatively weak KMT interaction with $\alpha=0.06$, which is half of the $\alpha_c$ or about one third
of the vacuum $\alpha$ given in \cite{Frank:2003ve} .

We still first study the chiral phase transition for a fixed weak coupling $g_v^s=0.2g_s$ by varying the $g_v^v$. 
The corresponding phase diagrams are shown in Fig.~\ref{fig:pdui30GV02}. We see that the separate first-order chiral 
transitions still appear for the small coupling $g_v^v=0.1g_s$. With rasing $g_v^v$, the two phase boundaries 
get closer and shorter. We notice that the left phase boundary shortens more significantly with $g_v^v$ compared to  
the right one. For the moderate coupling $g_v^v=0.4g_s$, the left phase boundary eventually vanishes   
(totally turns into crossover) but the right one still survives. So due to the vector interactions, the separation of 
the chiral transition can still be removed even the $\alpha$ is obviously less than the $\alpha_c$.

The investigation is then extended to a fixed moderate coupling $g_v^s=0.4g_s$, with both $\delta\mu$ and
$\alpha$ unchanged. We obtain the qualitatively similar phase diagrams for varied $g_v^v$ as displayed in
Fig.~\ref{fig:pdui30GV04}. We see that a rather weak coupling $g_v^v=0.45g_v^s$ is already 
strong enough to change the left phase boundary into crossover due to the raised $g_v^s$. It indicates 
that the $g_v^v$ does not need to be larger than the $g_v^s$ for the emergence of only one critical endpoint. 
Fig.~\ref{fig:pdui30GV04} also shows that the right phase boundary first shortens and then elongates
with $g_v^v$, but the left one always gets shorter with $g_v^v$ until it vanishes. Our further numerical
calculations suggest that the single phase boundary can even survive for very strong $g_v^v=1.5g_s$.

Similar to Figs.~\ref{fig:pdui05GVfixed}-\ref{fig:pdui10GVS0p4fixed}, Figs.~\ref{fig:pdui30GV02}-\ref{fig:pdui30GV04}
display that the two separate phase boundaries approach each other with the increase of $g_v^v$. As mentioned, this
is due to the decrease of $|\delta\mu'|$ with $g_v^v$. To illustrate this point, the $\mu$-dependence
of $|\delta\mu'|$ for varied $g_v^v$ at $T=10\MeV$ are shown in Fig.~\ref{fig:effuiGV04}, where the $g_v^s$ and 
$\delta\mu$ are the same as that in Fig.~\ref{fig:pdui30GV04}. We see that compared to the $|\delta\mu|$, the 
$|\delta\mu'|$ in between the left critical (or pseudo-critical) chemical potential and the right one reduces 
dramatically with $g_v^v$.

Distinct from Figs.~\ref{fig:pdui05GVfixed}-\ref{fig:pdui10GVS0p4fixed} (and Fig.~2 in \cite{Frank:2003ve}),
Figs.~\ref{fig:pdui30GV02}-\ref{fig:pdui30GV04} exhibit that with increasing $g_v^v$ it is the transition of one 
phase boundary into crossover rather than the coincidence of the two which results in only one
critical endpoint. This is because the left chiral transition for d quark is softened more significantly by both 
$g_v^v$ and $\alpha$. First, raising $g_v^v$ leads to the decrease (increase) of the $\mu_{d(u)}'$ according to 
Eq.~\eqref{eq:mixmu}. So the left chiral transition for d quark is weakened more significantly with $g_v^v$ 
compared to the right one for u quark. This point is also clearly demonstrated in Fig.~\ref{fig:MGV04}, where 
the quark masses as functions of $\mu$ at $T=10\MeV$ are plotted. Note that even $\mu_u'$ raises with $g_v^v$ 
for a fixed $g_v^s$, it does not mean that the right chiral transition for u quark must be intensified. Actually,
Figs.~\ref{fig:pdui30GV02}-\ref{fig:pdui30GV04} show that the right chiral transition is softened slightly with 
$g_v^v$ up to a moderate coupling strength (The reason will be given below).
Second, around the left phase boundary, the u quark condensate is still sizable and thus makes a relatively
large contribution to the d quark mass via the anomaly-related flavor-mixing.
In contrast, near the right phase boundary, the d quark condensate is suppressed significantly and thus its
contribution to the u quark mass is relatively small. This implies that the left chiral transition for d quark 
is also weakened more notably by the KMT interaction
\footnote{Here the contribution of u quark condensate to the d quark mass can be regarded as an effective
increase of the current mass of d quark, and vise versa. So the chiral phase transition for d quark is softened
by the KMT interaction.}.
This is why the left chiral boundary disappears (totally turns into crossover) but the right one still remains
for weak $\alpha$ and moderate (or weak) $g_v^v$.

Compared to the case with $\alpha=0$, Figs.~\ref{fig:pdui30GV02}-\ref{fig:pdui30GV04} demonstrate that it is
not the vector coupling difference but the strength of $g_v^v$ which is crucial for the appearance of a single
critical endpoint. Especially, even only one phase boundary appears in Figs.~\ref{fig:pdui30GV02}-\ref{fig:pdui30GV04}
for weak and moderate $g_v^v$, the d quark mass has reduced significantly near the left side of the true
phase transition, as displayed in Fig.~\ref{fig:MGV04}. This is different from what shown in
Fig.~2 of \cite{Frank:2003ve} for $\alpha>\alpha_c$, where both masses of u and d quarks drop suddenly
from large values to small ones across the phase boundary.

Let us explain why the right chiral transition for u quark is also weakened slightly with $g_v^v$  up to a
moderate strength. The reason can be attributed to the abrupt increase of $\rho_u$ and the relatively mild
change of $\rho_d$ near the right phase boundary. To understand this point, we can make a rough estimate of
the variation of $\mu_u'$ at the right critical chemical potential $\mu_c$ at $T=0$
(we use $\rho_f^{l(r)}$ and $\mu_f^{'l(r)}$ to denote the quark density and chemical potential on the left (right)
side of $\mu_c$, respectively). The effective u quark chemical potential on the left side of $\mu_c$ can be
approximated as $\mu_{u}^{'l}\approx{\mu_c+\mu_I-2g_v^s\rho_d^l+2g_v^v\rho_d^l}$ according to Eq.~\eqref{eq:shiftmu}
since $\rho_d^l\gg\rho_u^l\approx{0}$. Similarly, on the right side of $\mu_c$, we obtain
$\mu_{u}^{'r}\approx{\mu_c+\mu_I-2g_v^s\rho_d^l-2g_v^s\rho_d^l}$ using the approximations
$\rho_u^r\approx{\rho_d^r}\approx{\rho_d^l}$. These simplified expressions indicate that the $\mu_{u}^{'l}$ raises
with $g_v^v$ while the $\mu_{u}^{'r}$ remains the same. In general, on the same side of $\mu_c$, the larger
the $\mu_u'$, the smaller the u quark mass. So that the $\mu_{u}^{'l}$ increases evidently but the $\mu_{u}^{'r}$ 
keeps almost unchanged with $g_v^v$ imply that the abrupt drop of u quark mass at $\mu_c$ is weakened.

Note that such an explanation only holds for weak and moderate $g_v^v$, as exhibited in Fig.~\ref{fig:MGV04}. 
This is because the approximations ${\rho_d^r}\approx{\rho_d^l}$ and $\rho_d^l\gg\rho_u^l$ adopted above are 
no longer proper for the strong $g_v^v$ since near the $\mu_c$ the $m_{d}^{'l}$ becomes significantly larger 
than the $m_{d}^{'r}$ and the gap between the $m_{u}^{'l}$ and $m_{d}^{'l}$ also reduces obviously due to the 
decrease of $|\mu_I'|$. Actually, Fig.~\ref{fig:MGV04} shows that for the strong coupling
$g_v^v=0.8g_s$ the abrupt change of u quark mass across the phase boundary is not weakened but strengthened
compared to the case for $g_v^v=0.6g_s$, which is consistent with the decrease of $\mu_u'$ with $g_v^v$.

We stress that Fig.~\ref{fig:MGV04} also shows that the crossover chiral transition for d quark becomes less 
and less obvious with $g_v^v$, especially for $g_v^v>g_v^s$. The reason is that besides the influence of the 
$\alpha$ and the increased $g_v^v$, the crossover is also softened by the flavor-mixing due to the vector
coupling difference. We see that the chiral transition for the strong coupling $g_v^v=0.8g_s$ is already the type
displayed in Fig.~2 of \cite{Frank:2003ve} for $\alpha>\alpha_c$: Namely, the only phase boundary can still be 
regarded as the coincidence of the left and right first-order transition lines driven by the vector and KMT
interactions. Unlike the case for zero $\alpha$, the flavor-mixing due to the vector coupling difference plays 
a relatively minor role here even the chosen $\alpha$ is just about one third of its vacuum value.
Or in other words, the flavor-mixing due to vector interactions is unnecessary for the only phase boundary if the
$g_v^v$ is strong enough and the KMT interaction is not very weak.

In Ref.~\cite{Bratovic:2012qs}, it is argued that the ratio $g_v^s/g_s$ in a Polyakov-loop extended three-flavor NJL
model is likely to be larger than 0.4. As mentioned, our numerical study suggests that the $g_v^v$ obtained using
the same method is about 10\% larger than the $g_v^s$ according to the recent two-flavor lattice data. If such an
estimation is reliable, our model study suggests that the separate chiral transitions due to the isospin asymmetry may 
be still impossible in heavy ion collisions if the axial anomaly is suppressed effectively but not very significantly
near the phase boundary. Actually, Fig.~\ref{fig:pdui30GV04} shows that even the weak coupling strength $g_v^v=0.2g_s$ 
is already strong enough to change the left phase boundary for d quark to a rapid crossover for the moderate 
coupling $g_v^s=0.4g_s$
\footnote{This value locates in the range of $0.25g_s$ and $0.5g_s$, which are obtained from the instanton liquid
molecule model \cite{Schafer:1996wv} and the Fierz transformation of the one gluon exchange interaction, respectively}
. In addition, even it is proposed in \cite{Sasaki:2006ws} that ratio $g_v^v/g_v^s$ may locate in the range 1/3 and 1,
it is argued in \cite{Ferroni:2010xf} that this value rapidly approaches to 1 for $T>T_c$. All these arguments support
that the $g_v^v$ may not be very weak near the phase boundary, at least for small density. So even our choice of the 
$\alpha$ is just half of the $\alpha_c$ or one third of its vacuum value, the main conclusion for $\alpha>\alpha_c$ 
in \cite{Frank:2003ve} may still hold owning to the vector interactions.

All the above calculations are performed by changing the $g_v^v$ or $g_v^s$ with the $\alpha$ unchanged. Instead, we 
can do the same calculations by fixing the $g_v^v$ and $g_v^s$ and varying the $\alpha$. We then observe that the critical 
value of the $\alpha$ for the disappearance of the separation of the chiral transition reduces significantly compared to 
the $\alpha_c$ obtained in Ref.~\cite{Frank:2003ve} if the vector interactions are not very weak (especially the $g_v^v$). 
Of course, if both $g_v^s$ and $g_v^v$ are all strong, there is only the crossover transition no matter to what degree 
the axial anomaly is suppressed.

\subsection{ Validity of phase quenching in mean field NJL model with mismatched vector interactions }

Recently, the equivalence of QCD at finite $\mu$ and $\mu_I$ with a large number of colors $N_c$ has been
proposed~\cite{Hanada:2011ju,Hanada:2012es,Hidaka:2011jj}. The equivalence may enable the people to study the
properties of QCD at finite $\mu$ via the calculations of lattice QCD at finite $\mu_I$. The detailed discussion
on the validity of the phase quenching outside of the pion condensation region has been given in
Ref.~\cite{Hanada:2012es}, where the equivalence is confirmed in several popular QCD models at the MFA. Especially,
it is argued that the phase quenching still holds at the MFA of the NJL model even taking into account the 
flavor-mixing induced by instantons.

One evidence for the validity of phase quenching in \cite{Hanada:2012es} is that the free energies at finite
$\mu$ and $\mu_I$ are identical at the MFA in these QCD models, namely
\beq
\Omega_{M}(\mu_u=\mu_0,\mu_d=\mu_0,T)=\Omega_{M}(\mu_u=\mu_0,\mu_d=-\mu_0,T),\label{eq:phasequenching}
\eeq
where $\mu_0$ and $T$ are located in the region without pion condensation. In Ref.~\cite{Hanada:2012es}, such 
an equality is also obtained in the NJL model without considering the vector interactions. Here we stress that the
relation \eqref{eq:phasequenching} is still valid if vector interactions with the same couplings are included. 
This is because even the quark chemical potentials for u and d are modified, they are shifted by quantities with the
same magnitude due to the relations $\rho_B|_{\mu_I=0}=\rho_I|_{\mu=0}$ and $\rho_u|_{\mu_I=0}=-\rho_d|_{\mu=0}$
for $g_v^v=g_v^s$. So if the one-gluon exchange type interaction used in \cite{Hanada:2012es} is adopted by taking 
into account the vector channels, the phase quenching is still satisfied at the mean field level in the Hartree 
approximation. This is also consistent with the large-$N_c$ analysis given in \cite{Hanada:2012es}.

However, when considering the mismatched vector interactions, the phase quenching or the equivalence aforementioned
becomes invalid in this model even at the MFA. The reason is that for finite $\mu$ and zero $\mu_I$ and finite $\mu_I$ and
zero $\mu$, the effective baryon and isospin chemical potentials are modified by the couplings $g_v^s$ and $g_v^v$, respectively
(see Eq.~\eqref{eq:shiftmu}). Consequently, the effective chemical potential $\mu_{u}'|_{\mu_I=0}$ is no longer equal to
-$\mu_{d}'|_{\mu=0}$ and so the equality \eqref{eq:phasequenching} does not hold again according to Eq.~\eqref{eq:omega}. 

In this case, the degree of the equivalence breaking depends on the vector coupling difference. For instance, if $g_v^v=g_v^s/3$
as proposed in \cite{Sasaki:2006ws}, the equivalence in the MFA will be violated seriously. In addition, the one-gluon 
exchange type interaction in the Hartree-Fock approximation indicates that the vector coupling difference is subleading 
in $1/N_c$ according to Eq.~\eqref{eq:HatreeFock}. So for $N_c=3$, the phase quenching is also broken obviously for such 
an interaction in the MFA
\footnote{Note that the KMT interaction is also $1/N_c$ suppressed. However, the phase quenching is still exact at the MFA when
the KMT interaction is included, as argued in \cite{Hanada:2012es}.}.
On the other hand, the constraints from the chiral curvatures in the lattice QCD calculations suggests that vector coupling 
difference is not so large near $T_c$ at zero $\mu$ or $\mu_I$. Such an estimation suggests that the deviation of the 
equivalence at the MFA will be not so significant in the NJL model, at least for small chemical potentials.

It is also interesting to investigate the violation of the phase quenching at the MAF in other QCD models. In particular,
the similar study can be extended directly to the quark meson model \cite{Jungnickel:1995fp} of QCD by including the vector
interactions \cite{Ueda:2013sia}. If the quark-vector couplings $g_\omega$ and $g_\rho$ in \cite{Ueda:2013sia} are
different, the so called phase quenching at the MAF will be broken too in this model.

\section{ Discussion and conclusion }

We have studied the influence of vector interactions with different coupling constants in the isoscalar 
and isovector channels on the possible separation of the chiral transition under the isospin asymmetry 
in a two-flavor NJL model, where the $U(1)_A$ symmetry is assumed to be restored effectively near the 
phase boundary. In addition, the effect of the mismatched vector interactions on the proposed equivalence 
for the chiral transitions at finite $\mu$ and $\mu_I$ has also been studied at the MFA in this model.

We first show that, besides the argument based on the empirically different nucleon and vector-meson 
couplings \cite{Sasaki:2006ws}, the one-gluon exchange type interaction can also give rise to unequal 
vector interactions with $g_v^s>g_v^v$ at the MFA when including the Fock contribution. By extending 
the work \cite{Bazavov:2012qja} to finite $\mu_I$, we then obtain the quite different vector coupling 
difference with $g_v^s<g_v^v$ from the constraints of lattice chiral curvatures at zero/small quark 
chemical potentials. We demonstrate that, similar to the mass-mixing induced by the KMT interaction, 
the density-mixing due to the modified quark chemical potentials is produced owning to the mismatched 
vector interactions.

For the weak isospin asymmetry, we find that to convert the two separate chiral transitions into one, the $g_v^v$ 
must be significantly stronger than the $g_v^s$ without the axial anomaly. In this case, the non-anomaly flavor-mixing
induced by the vector interactions impacts the phase transition separation in the similar way as the anomaly one 
induced by instantons: the two detached phase boundaries get closer first and then coincide with the enhancement of 
the flavor-mixing. For the weak KMT interaction (the chosen coupling strength is about one third of the vacuum value) 
and relatively strong isospin asymmetry (the same as in \cite{Frank:2003ve}), we find that it is the strength of
 $g_v^v$ rather than the vector coupling difference which is crucial 
for the only single phase boundary. In particular, the separate chiral transitions disappear for the moderate or even 
weak $g_v^v$, not because of the overlap of the two phase boundaries, but because of the conversion of the left one into 
crossover. This is distinct from what found in \cite{Frank:2003ve} for $\alpha>\alpha_c$ without the vector interactions 
and the aforementioned coincidence of the phase boundaries without the KMT interaction. The reason is that under the 
isospin asymmetry, the left chiral transition for d quark is softened more significantly by both the vector-isovector 
and KMT interactions and eventually turns into crossover in advance.

Physically, the $g_v^v$ may not be much stronger than the $g_v^s$ near the phase boundary. So even the mismatched
vector interactions can lead to a non-anomaly flavor-mixing, its effect on the separation of the chiral transition is
limited unless the $|\mu_I|$ is very small. This seems to indicate that the two critical endpoints due to finite $\mu_I$
are still possible and may be observed in heavy ion collisions if the KMT interaction is very weak. However, we remark
that the effective restoration of the $U(1)_A$ symmetry obtained in recent lattice simulations does not imply
that the effect of axial anomaly can be ignored near $T_c$. Actually, the remnant $U(1)_A$ breaking around $T_c$ for zero
density is still observed in these studies. On the other hand, the constraints from the lattice chiral curvatures and flavor
susceptibilities all indicate that the strengths of $g_v^v$ and $g_v^s$ are considerable around $T_c$ for zero density
compared to the scalar interaction. We can expect that near the phase boundary at finite density,  the KMT interaction may
also not be very weak and the vector interactions are still appreciable. In this sense, the separate chiral transitions 
due to the isospin asymmetry could be still impossible in heavy ion collisions because of the vector interactions even
the instanton effect may be suppressed effectively.

We also revisit the validity of the phase quenching in the MFA of the NJL model by including the vector interaction. 
We first confirm that the equivalence for the chiral transition at finite $\mu$ and $\mu_I$ out of the pion condensation 
region proposed by Hanada et al. is still valid at the MFA for $g_v^v=g_v^s$. We then point out that such an equivalence
is broken explicitly by the mismatched vector interactions even at the MFA and the degree of this violation is dependent on
the vector coupling difference.

Note that recently the Polyakov-Loop extended NJL model has been extensively used to investigate the thermal and dense
properties of QCD. We stress that even our study is based on the NJL model, introducing the Polyakov-Loop dynamics does
not qualitatively change our main conclusions. In addition, our study can be directly extended to the quark meson model 
of QCD by incorporating the quark-vector-meson couplings and the axial anomaly. 

In this paper, we only study the chiral transition at relatively small $|\mu_I|$. In the neutron star 
core, the isospin asymmetry required by the charge neutrality and $\beta$-equilibrium may not be so weak. 
Especially, for $|\mu_I|>m_{\pi}/2$, the pion condensed matter may appear \cite{Son:2001,Kogut:2002,He:2005nk}. 
Moreover, the color superconductivity is also not considered in our calculation. The roles of the mismatched 
vector interactions in these research topics deserve further investigations.

\vspace{5pt}
\noindent{\textbf{\large{Acknowledgements}}}\vspace{5pt}\\
Z.Z. was partially supported by the NSFC ( No.11275069 ), by the Fundamental Research
Funds for the Central Universities of China, and by the University Plan of NCEPU for the Promotion
of Arts and Sciences.

\end{document}